\documentclass[aps, prd, amsmath, floats, floatfix, twocolumn,
superscriptaddress, nofootinbib, showpacs]{revtex4-1}

\usepackage{graphicx}
\usepackage{color}
\usepackage{soul}
\usepackage{url}
\usepackage{bm}         
\usepackage{times}
\usepackage{dcolumn}
\usepackage{bm}
\usepackage{epsf}
\usepackage{amssymb}

\newcommand{\beq}{\begin{equation}}
\newcommand{\eeq}{\end{equation}}
\newcommand{\beqn}{\begin{eqnarray}}
\newcommand{\eeqn}{\end{eqnarray}}

\usepackage{color}

\newcommand{\Caltech}{\affiliation{TAPIR, Walter Burke Institute for Theoretical Physics, MC 350-17,
    California Institute of Technology, Pasadena, California 91125, USA}}

\newcommand{\Cornell}{\affiliation{Cornell Center for Astrophysics and
    Planetary Science, Cornell University, Ithaca, New York, 14853, USA}}
\newcommand{\WSU}{\affiliation{Department of Physics \& Astronomy,
	Washington State University, Pullman, Washington 99164, USA}}
\newcommand{\CITA}{\affiliation{Canadian Institute for Theoretical 
    Astrophysics, University of Toronto, Toronto, Ontario M5S 3H8, Canada}}

\newcommand{\AEI}{\affiliation{Max Planck Institute for Gravitational Physics (Albert Einstein Institute), Am M\"uhlenberg 1, Potsdam 14476, Germany}}
\newcommand{\Maryland}{\affiliation{Department of Physics, University of Maryland, College Park, MD 20742, USA}}

\newcommand{\UNH}{\affiliation {Department of Physics, University of New Hampshire, 9 Library Way, Durham NH 03824, USA}}
\newcommand{\NCSA}{\affiliation{NCSA, University of Illinois at Urbana-Champaign, Urbana, Illinois, 61801, USA}} %

\newcommand{\GRAPPA}{\affiliation{GRAPPA, Anton Pannekoek Institute for Astronomy and Institute of High-Energy Physics, University of Amsterdam, Science Park 904, 1098 XH Amsterdam, The Netherlands}}
\newcommand{\DeltaITP}{\affiliation{Delta Institute for Theoretical Physics, Science Park 904, 1090 GL Amsterdam, The Netherlands}}
\newcommand{\Nikhef}{\affiliation{Nikhef, Science Park 105, 1098 XG Amsterdam, The Netherlands}}

\begin{document}

\title{Gravitational waveforms from SpEC simulations : neutron star-neutron star and low-mass black hole-neutron star binaries}

\author{F. Foucart}\UNH
\author{M.D. Duez}\WSU
\author{T. Hinderer}\GRAPPA\DeltaITP
\author{J. Caro}\WSU
\author{Andrew R. Williamson}\GRAPPA \Nikhef
\author{M. Boyle}\Cornell
\author{A. Buonanno}\AEI \Maryland
\author{R. Haas}\NCSA
\author{D.A. Hemberger}\Caltech
\author{L. E. Kidder}\Cornell
\author{H.P. Pfeiffer}\AEI\CITA
\author{M.A. Scheel}\Caltech

\begin{abstract}
Gravitational waveforms from numerical simulations are a critical tool to test and analytically calibrate the
waveform models used to study the properties of merging compact objects. In this paper,
we present a series of high-accuracy waveforms produced with the SpEC code for systems involving at least one neutron star.
We provide for the first time waveforms with sub-radian accuracy over more than twenty cycles for low-mass black hole-neutron star binaries, including binaries with
non-spinning objects, and binaries with rapidly spinning neutron stars that maximize the impact on the gravitational wave signal of the near-resonant growth of the fundamental excitation 
mode of the neutron star (f-mode). We also provide for the first time with SpEC a high-accuracy neutron star-neutron star waveform.
These waveforms are made publicly available as part of the SxS catalogue. We compare our results to analytical waveform models currently implemented in data analysis pipelines. For most simulations, the models lie outside of the predicted numerical errors in the last few orbits before merger, but do not show systematic deviations from the numerical results: comparing different models appears to provide reasonable estimates of the modeling errors. The sole exception is the equal-mass simulation using a rapidly counter-rotating neutron star to maximize the impact of the excitation of the f-mode, for which all models perform poorly. This is however expected, as even the single model that takes f-mode excitation into account ignores the significant impact of the neutron star spin on the f-mode excitation frequency.
\end{abstract}


\maketitle

\section{Introduction}
\label{sec:intro}

Numerical simulations of neutron star-neutron star (NSNS) and black hole-neutron star (BHNS) binaries play a critical role
in current efforts to model the gravitational wave (GW) and electromagnetic (EM) signals powered by these systems. The recent observation
of gravitational waves likely powered by a NSNS merger (GW170817), followed by signals across the entire EM spectrum,
confirmed that NSNS merger events have a non-negligible event rate~\cite{TheLIGOScientific:2017qsa,GBM:2017lvd,2017ApJ...848L..13A,TheVirgo:2014hva,TheLIGOScientific:2014jea}. GW170817 also allowed us to begin using NSNS mergers to study the
internal structure of neutron stars~\cite{Read:2008iy,DelPozzo:13,Lackey2014,GW170817-NSRadius,GW170817-PE}, the production of short-hard gamma-ray bursts~\cite{moch:93,Lee1999a,Janka1999,GBM:2017lvd,2017ApJ...848L..13A,2018Natur.561..355M}, and the synthesis of r-process elements~\cite{Li:1998bw,1976ApJ...210..549L,Rosswog:1998hy, 2005astro.ph.10256K,2010MNRAS.406.2650M,Metzger2017,2017Sci...358.1559K,2017Sci...358.1556C,2017ApJ...848L..19C,2017Sci...358.1556C,Cowperthwaite:2017dyu,2017Natur.551...80K,2017Sci...358.1583K,2017ApJ...848L..32M,2017ApJ...848L..18N,2017Natur.551...67P,2017Natur.551...75S,2017ApJ...848L..16S,2017ApJ...848L..27T,2017Sci...358.1565E}.
BHNS mergers, once detected, will allow us to study similar processes.

Placing constraints on the internal structure of neutron stars through GW observations requires us to model with sufficient 
accuracy the dependence of the GW signal on the parameters of the binary.
To first order, the effect of the finite size of neutron stars on the GW signal is set by the tidal deformability of the neutron stars, $\Lambda =2/3 k_2 (R_{\rm NS}c^2/[GM_{\rm NS}])^5$ 
-- or, more accurately, by the effective tidal deformability $\tilde \Lambda$ of the binary, a linear combination of the $\Lambda$ of the merging compact objects~\cite{Flanagan2008,Read:2008iy,Lackey2014}. 
Here, $k_2$ is the Love number of the neutron star, $R_{\rm NS}$ its radius, and $M_{\rm NS}$ its mass. $G$ is the gravitational constant and $c$ is the speed of light. $\Lambda$ is thus mainly sensitive to
the compaction of the neutron star, $C_{\rm NS} = GM_{\rm NS}/(R_{\rm NS}c^2)$. GW170817 alone provided interesting constraints on $\Lambda$~\cite{GW170817-PE}, and better results are expected once information from
multiple merger events can be combined. 

An important role of numerical simulations in the era of GW astrophysics is to provide reliable templates for the GWs produced by
a given binary merger. General-relativistic hydrodynamics simulations of NSNS and BHNS mergers have steadily improved the accuracy of their GW predictions
since the first general relativistic simulations of these systems~\cite{Shibata01,2007CQGra..24S.125S}. Due to the need to evolve the neutron star matter, NSNS and BHNS simulations are typically orders of magnitude less accurate
than binary black hole (BBH) simulations, and until recently simulations were unable to more than marginally resolve finite-size effects in the GW signal. An important advance towards high-
accuracy waveforms was the implementation of high-order numerical methods for merger simulations~\cite{Radice:2012cu,Radice:2013cba}. A combination of high-order methods and/or improved mesh refinement
algorithm has allowed multiple groups to provide numerical GW templates with sub-radian accuracy over more than 10 orbits for NSNS binaries~\cite{2017PhRvD..96h4060K,2018arXiv180601625D}, an important threshold considering that finite-size effects typically lead to the accumulation of a few radians of dephasing between a NSNS/BHNS system and an equivalent BBH system. For BHNS binaries, modern studies have focused mostly on the characterization of the merger signal~\cite{Pannarale:2015jia,Kawaguchi2015}. No long, high-accuracy templates have been made available so far -- although some of the long BHNS simulations presented in this work were already used to test analytical models~\cite{Hinderer:2016eia}, and to study the impact of model uncertainties on our ability to measure $\Lambda$~\cite{2018arXiv180904349C}.

Numerical simulations of NSNS/BHNS mergers cannot be used directly for parameter estimation (PE) in the analysis of GW signals. PE studies require the
production of thousands of simulated GW signals, while a single merger simulation takes weeks to months to complete. Additionally, numerical simulations of compact binaries
are relatively short ($\lesssim 0.1\,s$), while PE studies require minutes-long templates. Accordingly, analytical and phenomenological models have been developed to capture both 
the inspiral phase (using analytical methods) and the merger phase (using either effective-one-body methods calibrated to BBH simulations, or phenomenological fits). Numerical simulations play a dual role in the study of GW signals from NSNS/BHNS binaries: they allow us to test the accuracy of existing models, and they give us the data necessary to calibrate improved models when these errors are found to be unacceptably large -- either due to improvements in the sensitivity of the detectors, or because we need models in a so-far unexplored part of parameter space.

The simulations presented in this paper are part of this community-wide effort to produce reliable numerical waveforms, and to use them to test and improve analytical models. We 
present a set of GW signals generated using the SpEC code~\cite{SpECwebsite}. All of our simulations have in common the use of high-order methods and very simple equations of state for the evolution of the neutron star matter, and most are meant for high-accuracy comparisons between analytical and numerical waveforms. They are also all performed at 3 distinct numerical resolutions. The numerical waveforms are made publicly available as part of the SxS catalogue of waveforms~\cite{SXSCatalog}, or through their respective DOIs~\cite{BHNSq1s0,BHNSq1s2m,BHNSq1.5s0,BHNSq2s0,BHNSq2s2m,BHNSq6s0,NSNSq1G2,NSNSq1MS1b}. We present 2 high-accuracy, $12-13$ orbits long BHNS simulations with low-mass, non-spinning black holes (mass ratios $q=M_{\rm BH}/M_{\rm NS}=\{1,2\}$), as well as a longer (and consequently less accurate) simulation of a mass ratio $q=1.5$ BHNS system. At more that $16.5$ orbits of evolution, this is the longest BHNS simulation produced to date. We also present the first high-accuracy simulations of BHNS binaries with {\it spinning} neutron stars: two simulations with mass ratios $q=\{1,2\}$, non-spinning black holes, and neutron stars with dimensionless spin $\chi_{\rm NS}=0.2$ anti-aligned with the orbital angular momentum. For spinning neutron stars, the equal-mass system is particularly interesting because the orbital frequency at which the f-mode of the neutron star comes into resonance with the orbital motion is low enough that dynamical tides are enhanced, and the binary inspiral is strongly accelerated. Finally, we also include 2 waveforms for NSNS binary mergers, which complement a number of high-accuracy NSNS waveforms already available in the literature.

The available configurations and our numerical methods are presented in Section~\ref{sec:methods}, and conservative error estimates for each simulation are discussed in Section~\ref{sec:errors}. We put these errors into context by comparing NSNS, BHNS, and BBH waveforms, thus estimating the magnitude of finite size effects in the chosen binary systems, in Section~\ref{sec:NRwaves}. Finally, we provide direct comparisons between our waveforms and a sample of the most advanced models for NSNS and BHNS waveforms existing today in Section~\ref{sec:models}.

\section{Methods}
\label{sec:methods}

\subsection{Initial Data}
\label{sec:ID}
For the majority of the systems evolved in this study, we generate constraint satisfying initial data using our in-house solver, Spells~\cite{Pfeiffer2003,Pfeiffer2003a}. Initially developed to generate initial data for black hole binaries, Spells was later adapted to BHNS binaries~\cite{FoucartEtAl:2008}, NSNS binaries~\cite{Haas:2016}, and the production of initial data for neutron stars of arbitrary spins~\cite{Tacik:2015tja,Tacik:2016zal}. The iterative algorithm used to generate initial data for BHNS and NSNS binaries is strongly inspired from the earlier work of
Gourgoulhon et al.~\cite{GourgoulhonEtAl2001a} and Taniguchi et al.~\cite{TaniguchiEtAl:2006}. All binaries generated with Spells have their orbital eccentricity reduced to $e\lesssim 0.002$ using the iterative method developed by Pfeiffer et al.~\cite{Pfeiffer-Brown-etal:2007}, with the exception of the shorter BHNS simulation with mass ratio $3$, which has $e\sim 0.008$ (eccentricity reduction is more difficult for binaries with small initial separation). A list of all initial configurations is
presented in Table~\ref{tab:ID}. 

\begin{table}
\begin{center}
\caption{Initial parameters of the binaries studied in this paper. $M_{1,2}$ are the masses of the objects, either the Christodoulou mass of 
the black hole or the ADM mass of an isolated non-spinning neutron star with the same equation of state and baryon mass as the neutron star under
consideration\footnote{For spinning neutron stars, we also considered defining $M_2$ as the mass of an isolated neutron star with the same baryon mass and spin as the simulated neutron star, leading to $M_2=1.40176$ for the spinning neutron stars in this paper. The phase difference with analytical model changes by less than $0.1{\rm rad}$ at merger between these two definitions, well below our numerical error for spinning neutron stars.}. By convention, $M_1 \ge M_2$, and $M_1$ is the black hole for equal mass BHNS systems. $\chi_{1,2}$ are the dimensionless spins of the objects, $N_{\rm cycles}$ is the number of cycles up to the maximum amplitude of the gravitational
wave signal, $\Omega_0$ is the initial angular velocity, and $M=M_1+M_2$ the total mass. Binary neutron star systems have names starting with NSNS, and black hole-neutron star systems have names starting with BHNS. EoS is the equation of state of the neutron star(s), described in more detail in the text.}
{
\begin{tabular}{|c||c|c|c|c|c|c|c|}
\hline
Model & $M_{1}\,(M_\odot)$ & $M_{2}\,(M_\odot)$ & $\chi_{1}$ & $\chi_{2}$& $N_{\rm cycles}$ & $\Omega_0 M$ & EoS \\
\hline
BHNSq1s0 & 1.4 & 1.4 & 0 & 0 & 24.5 &0.0175 & $\Gamma 2$ \\
BHNSq1s2m & 1.4 & 1.4 & 0 & -0.2 & 21.6 & 0.0175 & $\Gamma 2$ \\
BHNSq1.5s0 & 2.1 & 1.4 & 0 & 0 & 33.2 & 0.0158 & $\Gamma 2$ \\
BHNSq2s0 & 2.8 & 1.4 & 0 & 0 & 26.1 & 0.0187 & $\Gamma 2$\\
BHNSq2s2m & 2.8 & 1.4 & 0 & -0.2 & 24.7 & 0.0187 & $\Gamma 2$\\
BHNSq3s0 & 1.35 & 4.05 & 0 & 0 & 12.3 & 0.0285 & H1\\
NSNSq1$\Gamma$2 & 1.4 & 1.4 & 0 & 0 & 25.2 & 0.0165 & $\Gamma 2$ \\
NSNSq1MS1b & 1.35 & 1.35 & 0 & 0 & 16.4 &0.0192 & MS1b\\
\hline
\end{tabular}
\label{tab:ID}
}
\end{center}
\end{table}

Most of these initial conditions are chosen to maximize finite size and spin effects and minimize numerical errors, thus allowing the use of our waveforms for finer testing of analytical models. This is why we choose systems that are physically unlikely: an equal mass BHNS systems or a neutron star with $\chi=0.2$ are not expected to be observed. These considerations also drive our choice of equation of state: we choose an ideal gas equation of state with polytropic index $\Gamma=2$. The pressure is $P=101.45 \rho^\Gamma$ and the internal energy $u=(\Gamma-1) P$. With these parameters, a $1.4M_\odot$ neutron star has a large dimensionless tidal deformability $\Lambda=791$, at the upper end of what is currently allowed by constraints from gravitational wave observations~\cite{TheLIGOScientific:2017qsa}. The properties of the neutron stars evolved for the studies in this manuscript are summarized in Table~\ref{tab:EoS}. Equations of state providing better agreement with nuclear theory are of course available, and would certainly lead to different evolution of the post-merger remnant. However, nuclear-theory based equations of state cannot be evolved with as much accuracy. Most of the tidal models currently used to produce gravitational wave templates parametrize neutron stars solely through $\Lambda$, and the waveforms presented here allow for tests of these single-parameter models\footnote{Some Effective-One-Body models include the impact of the octupole, f-mode frequencies for quadrupole and octupole, and the spin-induced quadrupole, and thus in principle depend on multiple parameters. In current practical data analysis applications, quasi-universal relations are however used to reduce everything to the single $\Lambda$ parameter, and this was also done for the model waveforms used in this paper}. While studies have shown that $\Lambda$ is the most important parameter to model tidal effects~\cite{Lackey2011,ReadEtAl2013,Bernuzzi:2014kca}, it is likely that higher-accuracy numerical waveforms will eventually begin to capture corrections to the waveforms that do not solely depend on $\Lambda$. Dedicated studies comparing systems with the same $\Lambda$ but different equations of state will be necessary to determine the importance of these corrections.

\begin{table}
\begin{center}
\caption{Properties of the neutron stars used in this study. EoS is the name of the equation of state, $M_{\rm ADM}$ the ADM mass of the star
in isolation, $M_b$ its baryonic mass, $C=GM/Rc^2$ its compaction, and $\Lambda$ its dimensionless tidal deformability.}
{
\begin{tabular}{|c||c|c|c|c|}
\hline
EoS & $M_{\rm ADM}\,(M_\odot)$ & $M_{b}\,(M_\odot)$ & $C$ & $\Lambda$ \\
\hline
$\Gamma 2$ & 1.40 & 1.51 & 0.144 & 791\\
MS1b & 1.35 & 1.47 & 0.142 & 1540\\
H & 1.35 & 1.48 & 0.162 & 624\\
\hline
\end{tabular}
\label{tab:EoS}
}
\end{center}
\end{table}

We also present one NSNS and one BHNS waveform using a piecewise polytropic equation of state calibrated to a nuclear-theory model for cold dense matter (MS1b and H1~\cite{Read:2008iy}). These equations of state are complemented with a $\Gamma$-law thermal component. For the NSNS binary, we consider an equal mass, non spinning system and the MS1b equation of state. This waveform was generated as part of a code-comparison project, and to guarantee exactly identical initial data we use initial conditions produced using the LORENE code~\cite{LORENE,MS1bLORENEID}. The MS1b equation of state models unrealistically large stars (ruled out by GW observations). This simulation has larger constraint violations at $t=0$ than the polytropes, and the evolutions themselves are significantly less accurate -- in part because the MS1b equation of state is not as smooth as the $\Gamma$-law equation of state, and also possibly because of the necessity to use a wider grid spacing for such large neutron stars. The BHNS binary uses a mass ratio $q=3$ and the H1 equation of state, with initial data generated with Spells. It is a shorter simulation generated for the purpose of comparison with a similar configuration studied with the SACRA code~\cite{Kyutoku:2010zd}.
Error estimates for all of these binaries are discussed in Sec.~\ref{sec:errors}.

\subsection{Evolution Algorithm}

The initial conditions presented in Sec.~\ref{sec:ID} are evolved with the SpEC code~\cite{SpECwebsite}. SpEC evolves Einstein's equations of general relativity on a pseudo-spectral grid in the generalized harmonic formulation~\cite{Lindblom:2007}, with damped harmonic gauge conditions~\cite{Szilagyi:2009qz}. The general relativistic equations of hydrodynamics are evolved on a separate grid~\cite{Duez:2008rb} using fifth-order finite difference methods (MP5 reconstruction), as proposed by Radice et al.~\cite{Radice:2012cu}. Both systems of equations are evolved in time using third-order Runge-Kutta time stepping and identical time steps chosen adaptively to reach a target time discretization error. Source terms are communicated between the two grids at the end of each full Runge-Kutta step. Values of the source terms at intermediate times are obtained through linear extrapolation from the values stored at the end of the last two time steps. We refer the interested reader to~\cite{Duez:2008rb,Foucart:2013a} for a more detailed description of our algorithm. This mixture of numerical methods has both advantages and disadvantages. On the one hand, SpEC is generally capable to obtain high-accuracy waveforms at a fairly low computational cost: the longest $q=1.5$ simulation cost $(18,38,90)$kCPU-hrs from the beginning of the simulation to the peak of the gravitational waveform, at our 3 chosen resolutions on the zwicky cluster at Caltech\footnote{Simulations involving spinning neutron stars, piecewise polytropic equations of state, or with a tighter control of the amount of matter remaining on the grid can be up to $3-4$ times more expensive, while the shorter BHNS simulations with non-spinning neutron stars presented here are cheaper.}. On the other hand, as different parts of the code have different orders of convergence, errors of different signs, and may dominate the error budget at different times, measuring errors is a complex task. In Sec.~\ref{sec:errors}, we present different sources of errors and a conservative method to estimate the phase error in SpEC. In practice, we find that this estimate is often overly pessimistic, but prefer a cautious approach when presenting waveforms aimed mainly at calibrating analytical models.

\subsection{Numerical Setup}

Each of the cases discussed here is evolved at three different resolutions. The older simulations, for non-spinning BHNS binaries, use initial resolutions on the finite difference grid of $\Delta x = (329,263,220,188)\,{\rm m}$ (the $q=2$ case was not run at the highest resolution, the other cases were not run at the coarsest resolution), within a cubic box of initial length $L=26.3{\rm km}$.
\footnote{Our initial data for the neutron stars uses a conformally flat metric, leading to a coordinate radius significantly smaller than the circular
radius quoted in Sec.~\ref{sec:ID}, e.g. the $\Gamma 2$ neutron stars have a circular radius $R=14.4\,{\rm km}$ but a coordinate radius $R=11.5{\rm km}$} 
In SpEC, the numerical grid moves with the compact objects, and is in particular rotated and rescaled as they orbit and spiral in. This slowly increases the resolution of the grid in the lab frame, but also causes the size of the neutron star on the grid to grow. To counteract this effect, we regularly rescale the finite difference grid, interpolating the evolved variables onto a new, coarser grid when the binary inspirals. This approximately maintains a constant resolution in the inertial frame.

The BHNS binaries with spinning neutron stars use $\Delta x = (294,235,196){\rm m}$. They also use a more efficient grid construction algorithm: only regions in which matter is present are covered by the grid, and the code adaptively adds/removes small cubic blocks to the grid as needed to follow the fluid. As the grid still contracts when the binary inspirals, we interpolate onto a new grid matching the initial grid spacing in the inertial frame every time the resolution increases by $20\%$. The $\Gamma 2$ NSNS binary uses the same adaptive grid as the BHNS simulations with spinning NSs. For the MS1b NSNS binary, to match the prescriptions of the code comparison project, we use the coarser grid resolution $\Delta x = (368,294,235){\rm m}$.

The spectral grid uses adaptive refinement to automatically add/remove basis functions in each patch of the grid in order to obtain a target relative accuracy in the spectral expansion of the metric variables and of their spatial derivative. At the middle resolution, that target is $10^{-4}$ in the wave zone, and $10^{-8}$ close to the compact objects. The target accuracy is varied as $(\Delta x)^5$, with $\Delta x$ the resolution of the finite difference grid. The same method is used to choose the target accuracy of the adaptive time stepping algorithm, but with the middle resolution targeting a relative error of $10^{-4}$ and an absolute error of $10^{-6}$ in each of the evolved variable (see~\cite{Foucart:2013a} for details).

The merger and post-merger evolution methods are largely unchanged from our previous simulations~\cite{Foucart:2013a}, except for the use of the new adaptive finite difference grid. Once we have evolved the simulation for a few milliseconds past merger (defined as the time at which the amplitude of the GW signal peaks), we rapidly extract the gravitational waves by evolving Einstein's equations with no matter source terms. This clearly create large errors where the compact objects were located (especially for NSNS binaries), but these errors do not propagate faster than the speed of light, and thus do not affect the gravitational wave produced earlier in the simulation. This significantly reduces the cost of our simulations. For more realistic equations of state, following the post-merger evolution is of course interesting in itself. But when using idealized $\Gamma 2$ equations of state, no magnetic fields, and no neutrinos, as in the simulations presented here, it would be rather pointless to spend computational resources on a post-merger evolution that is largely unphysical.

\section{Error estimates}
\label{sec:errors}

The main intended use of the waveforms presented in this manuscript is to help calibrate semi-analytical waveform models. To avoid overfitting these models to numerical noise, we make the choice to construct conservative error estimates which likely overestimate numerical errors. We consider three main sources of errors. The most important is the error due to the spatial and time discretization of the problem. With the methods used in SpEC, we expect better than second order convergence from all sources of discretization errors (and we indeed observe such convergence on simpler problems when the numerical grids are static). However, multiple sources of errors enter our error budget: time discretization error, spatial discretization error on the spectral grid used to evolve Einstein's equations, spatial discretization error on the finite difference grid used to evolve the equations of hydrodynamics, interpolation error in the communication of source terms between the two grids, and extrapolation error for the determination of the source terms at intermediate time steps. These errors may be of the same order of magnitude, especially as the simulation parameters are chosen to avoid wasting resources by, e.g., taking extremely small time steps or pursuing significantly smaller errors on the spectral grid than on the finite volume grid. Additionally, the adaptive mesh refinement algorithm used on the spectral grid is a powerful tool to efficiently allocate computational resources, but it also modifies the grid at different times for different simulations, making standard convergence tests difficult. As a consequence, the phase difference between the waveforms generated at different resolutions can occasionally be very small despite non-negligible discretization errors. To obtain reliable error estimates, we perform each simulations with three different grid resolutions. Some simulations (BHNSq2s0, BHNSq1.5s0, BHNSq1s0) were additionally performed with multiple numerical algorithms (gauge choices, second-order accurate fluid evolution instead of fifth-order accurate fluid evolution) to verify that error estimates obtained with one algorithm are consistent with the results obtained for the same simulation but using a different algorithm. 

We compute the discretization errors as follow. Given a pair of simulations at different resolutions, we estimate the difference between the highest of the two resolutions and a theoretical infinite-resolution simulation using Richardson extrapolation of the error, assuming (pessimistically) second order convergence\footnote{We use the resolution of the finite difference grid for this calculation, as the tolerances of the spectral adaptive mesh refinement and of the adaptive time stepper are both tied to the resolution of the finite difference grid.}. We compute two error estimates in this manner, by comparing the highest resolution available to us with each of the other two resolutions separately. To avoid small error estimates due to cancellation of phase errors of opposite signs (typically due to different sign for the phase errors in the early and late inspiral), we then define our discretization error, $\Delta \phi_{\rm dis}$, as the worst of these two estimates.

We also include in our error calculations two effects that are generally smaller than the discretization error: the effect of mass loss at the boundary of the finite difference grid, and the error due to extrapolation of the gravitational wave signal to infinity from measurements made at finite radii. For the former, we estimate $\Delta \phi_{\rm dM} = (\delta M_{\rm NS}/M^b_{\rm NS}) \omega_{22} t$, following~\cite{Boyle2007}. Here $\delta M_{\rm NS}$ is the baryon mass lost by the NS(s) during inspiral, and $M^b_{\rm NS}$ the total mass of the NSs. We note that this conservatively assumes that all mass losses happen around $t=0$, causing maximal impact on the waveform, even though the observed mass losses are distributed over the entire simulation (and are in fact slightly larger at later times). For the latter, we compute the phase difference between waveforms extrapolated to infinity by fitting second and third order polynomials in $(1/R)$ to measurements at 20 radii equally spaced in $(1/R)$ between $100M$ and $450M$, with $M$ the total mass of the system. The extrapolation error $\Delta \phi_{\rm ext}$ is taken to be the maximum value of that phase difference for $t\in [0,t_{\rm merger}]$. Typically, $\Delta \phi_{\rm ext}\sim (0.01-0.05)\,{\rm rad}$ is the dominant source of error at early times but becomes negligible as we approach merger. The mass loss error tends to be much smaller than the discretization error, except for the equal mass, non-spinning BHNS binary.\footnote{Simulation BHNSq1s0 allowed more mass to leave the grid before requesting an expansion of the finite difference grid than other simulations, and additionally is the simulation with the smallest discretization error.} We estimate the total simulation error as
\beq
\Delta \phi_T = \sqrt{\Delta\phi_{\rm dis}^2 + \Delta\phi_{\rm ext}^2 + \Delta\phi_{\rm dM}^2}.
\label{eq:phiT}
\eeq

\begin{figure*}
\begin{center}
\includegraphics[width=.49\textwidth]{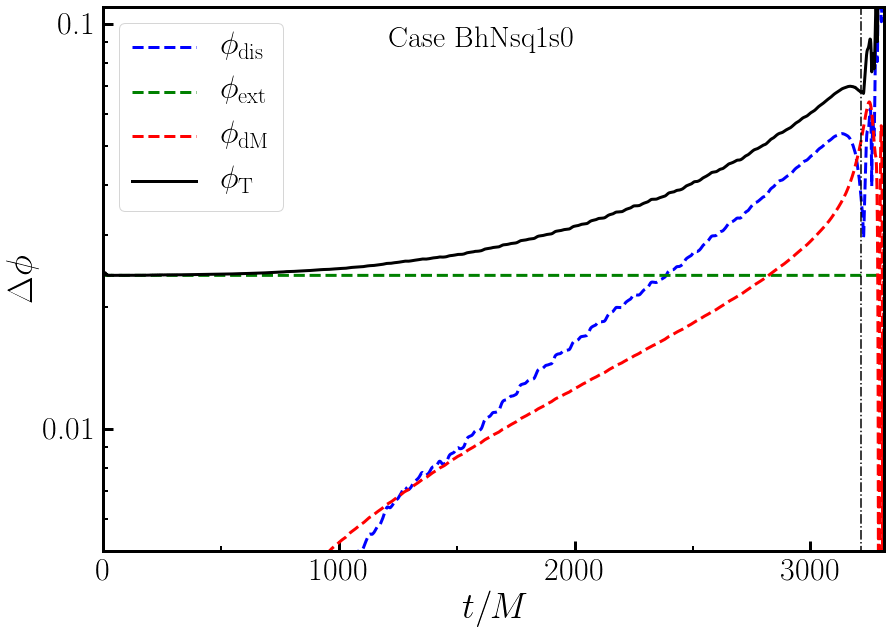}
\includegraphics[width=.49\textwidth]{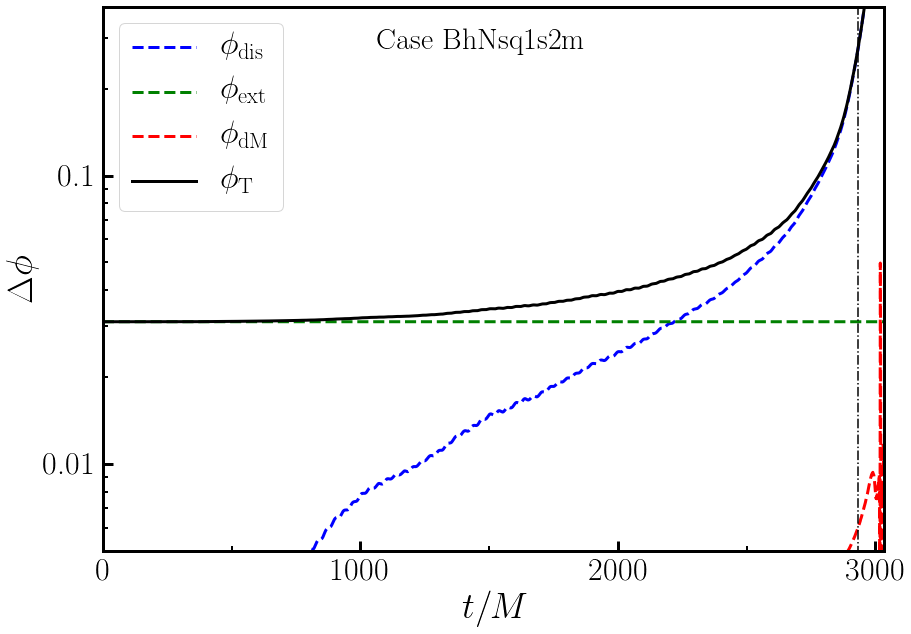}\\
\includegraphics[width=.49\textwidth]{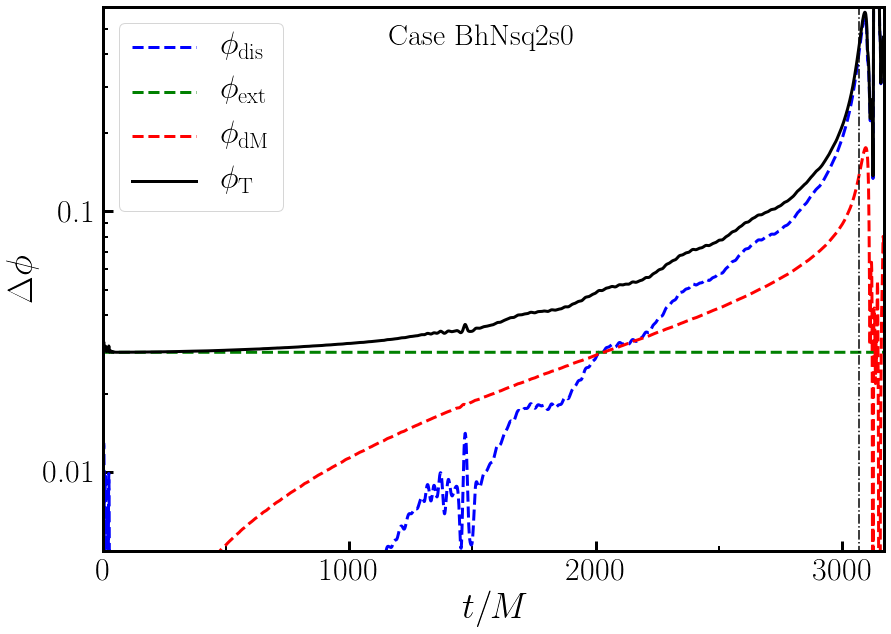}
\includegraphics[width=.49\textwidth]{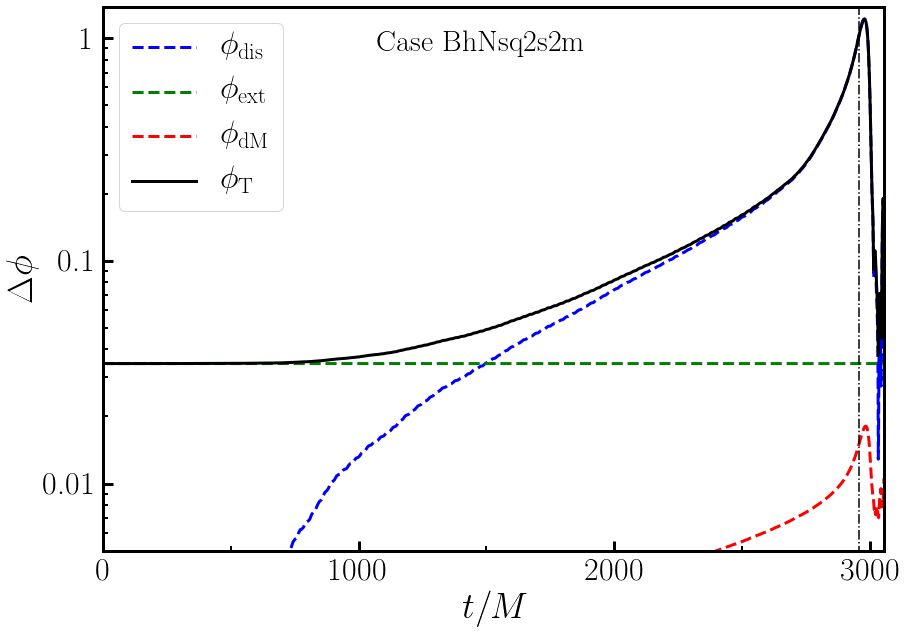}\\
\includegraphics[width=.49\textwidth]{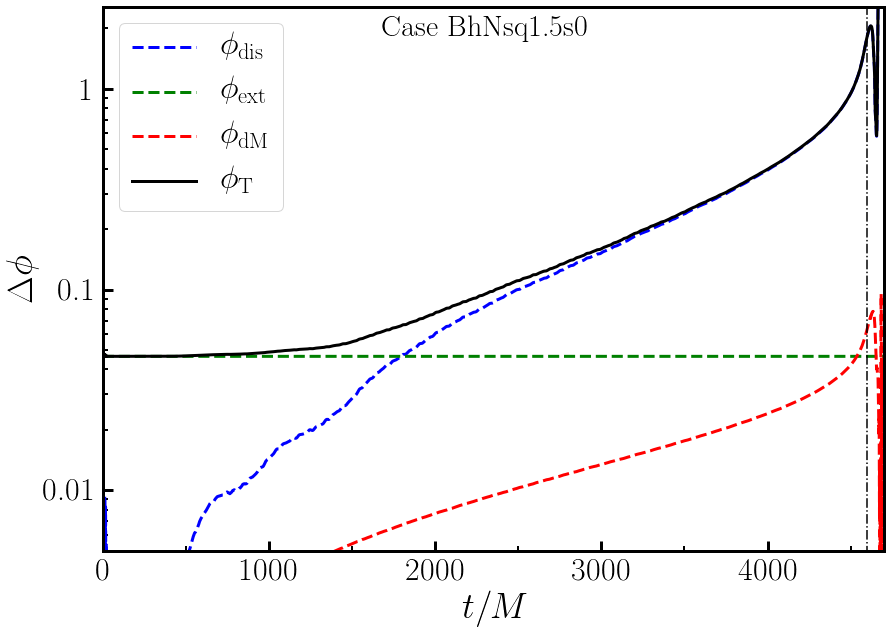}
\includegraphics[width=.49\textwidth]{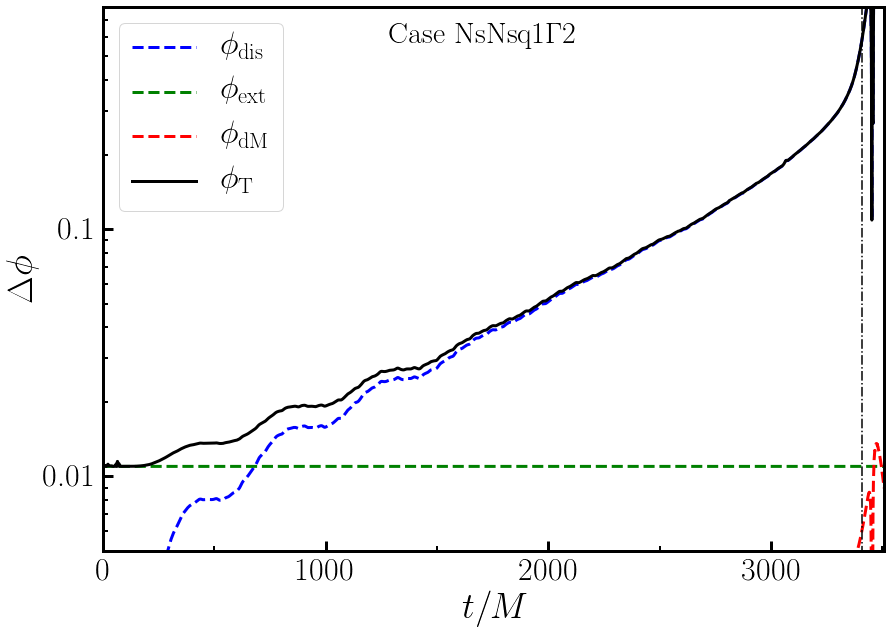}
\caption{Numerical error in the phase of the $(2,2)$ mode of the GW signal for the 6 simulations using a $\Gamma$-law equation of state. For each simulation, we show estimates of the discretization error (dashed blue), mass loss error (dashed red) and extrapolation error (dashed green), as well as the total numerical error (solid black line) defined by Eq.~(\ref{eq:phiT}). The vertical dashed line shows the time of maximum amplitude of the waveform.}
\label{fig:NumericalErrors}
\end{center}
\end{figure*}

\begin{figure}
\begin{center}
\includegraphics[width=.99\columnwidth]{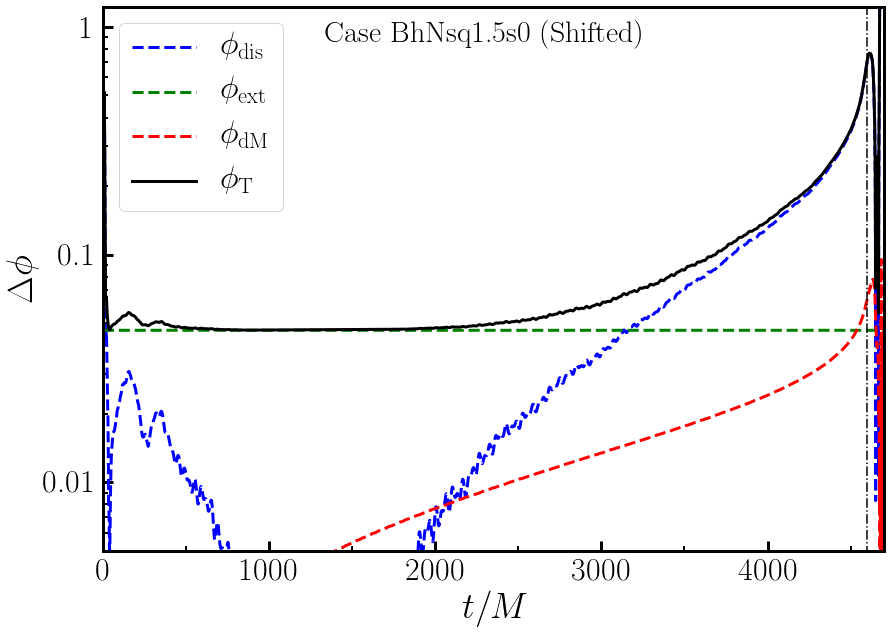}
\caption{Same as Fig~\ref{fig:NumericalErrors}, but after allowing for an arbitrary time and phase shift in the low-resolution results of case BHNSq1.5s0, minimizing phase errors in the time interval $[1000,1700]$.}
\label{fig:MatchedError15}
\end{center}
\end{figure}

The resulting error estimates for the dominant $(2,2)$ mode of the gravitational waveforms are shown in Fig.~\ref{fig:NumericalErrors}. Numerical errors are larger for $q=2$ than $q=1$, and larger for spinning binaries than for non-spinning binaries. The largest error is observed in the $q=1.5$ BHNS simulation, but this is simply a result of a significantly longer evolution time. 

Most of the error is due to small time offsets between resolutions incurred during the early evolution. That time offset is irrelevant when comparing numerical waveforms to analytical models, as the waveforms have to be matched through an arbitrary time and phase shift. When comparing numerical waveforms to analytical models, we compute errors in the same way, except that we allow for a time and phase shift of the waveform minimizing the root-mean-square phase difference in an interval $[t_{\rm min},t_{\rm max}]$. The result of this procedure for the $q=1.5$ simulation is shown in Fig.~\ref{fig:MatchedError15}. For that figure, we choose the end of the matching interval so that the time between $t_{\rm max}$ and the peak of the GW signal is comparable to the evolution time of the $q=1,2$ BHNS simulations. The phase error at merger is then reduced by more than a factor of $2$, and comparable to the $q=2$ results.

In the following sections, when matching simulations with different initial conditions or when matching simulations and analytical models, we will use this last method to compute numerical errors. However, the reader interested in the `raw' numerical errors, estimated without any time or phase shift, can refer back to Fig.~\ref{fig:NumericalErrors}.

\begin{figure}
\begin{center}
\includegraphics[width=.99\columnwidth]{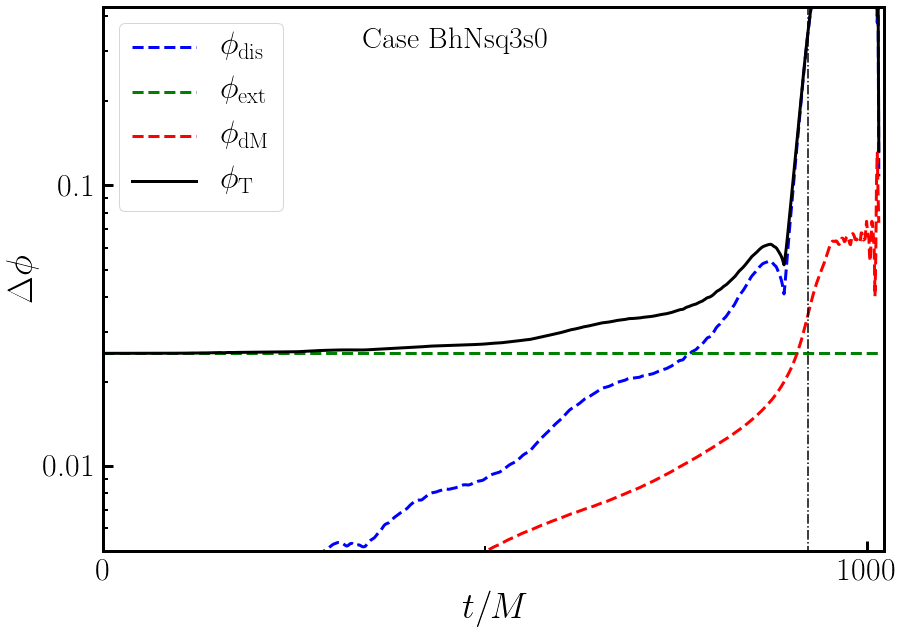}
\caption{Same as Fig~\ref{fig:NumericalErrors}, but the BHNS binary using the H1 equation of state.}
\label{fig:BHNSQ3}
\end{center}
\end{figure}

Fig.~\ref{fig:BHNSQ3} shows error estimates for the $q=3$ BHNS simulation with piecewise-polytropic (H) equation of state. The phase error at merger is small ($\Delta \phi \sim 0.3\,{\rm rad}$), though this is in part due to the shorter evolution time.

\begin{figure}
\begin{center}
\includegraphics[width=.99\columnwidth]{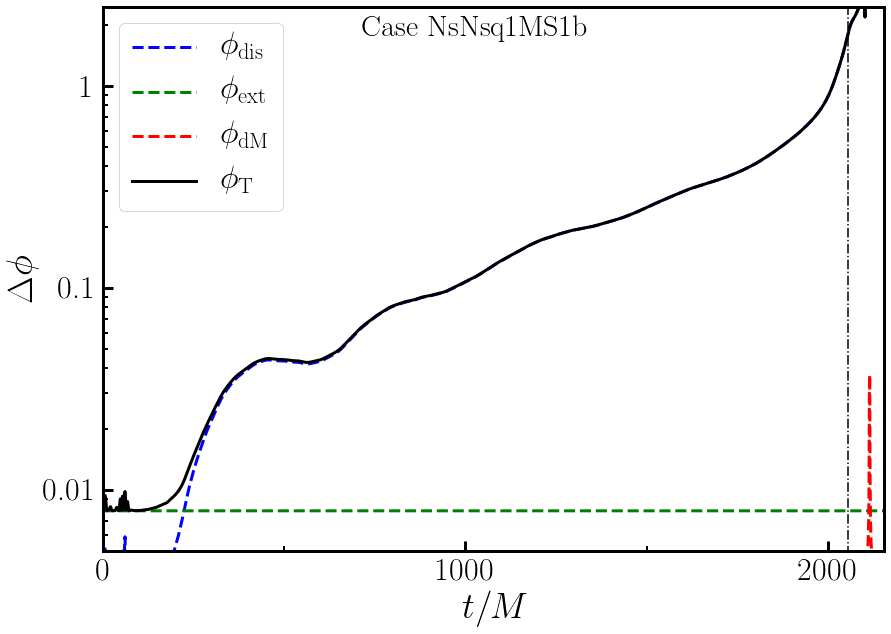}
\caption{Same as Fig~\ref{fig:NumericalErrors}, but for the NSNS simulation using the MS1b equation of state. In this case, the numerical error is nearly entirely due to the effect of unresolved transients at early times.}
\label{fig:ErrorMS1b}
\end{center}
\end{figure}

Finally, Fig.~\ref{fig:ErrorMS1b} shows error estimates for the NSNS simulation with MS1b equation of state. The effect of less accurate initial data and/or initial data interpolation error is obvious here: at early times, numerical errors are much larger here than in any other simulation, and so is the error at merger, despite the fact that the simulation itself is shorter. A time and phase shift may help reduce that error, but given the length of the simulation, this would leave only a small number of usable orbits. We should note that this is not an indication that LORENE data is less accurate than initial data generated with our own Spells solver. Instead, we argue that this is a general issue with initial data solvers using spectral methods -- as both Spells and LORENE do. The MS1b equation of state is not smooth, and this leads to larger errors in the spectral representation of the initial data. We have performed short simulations of neutron star mergers using piecewise-polytropic equations of state from Spells initial data, and find early time errors comparable to what is shown in Fig.~\ref{fig:ErrorMS1b}.

\section{Numerical waveforms}
\label{sec:NRwaves}

\begin{figure}
\begin{center}
\includegraphics[width=.99\columnwidth]{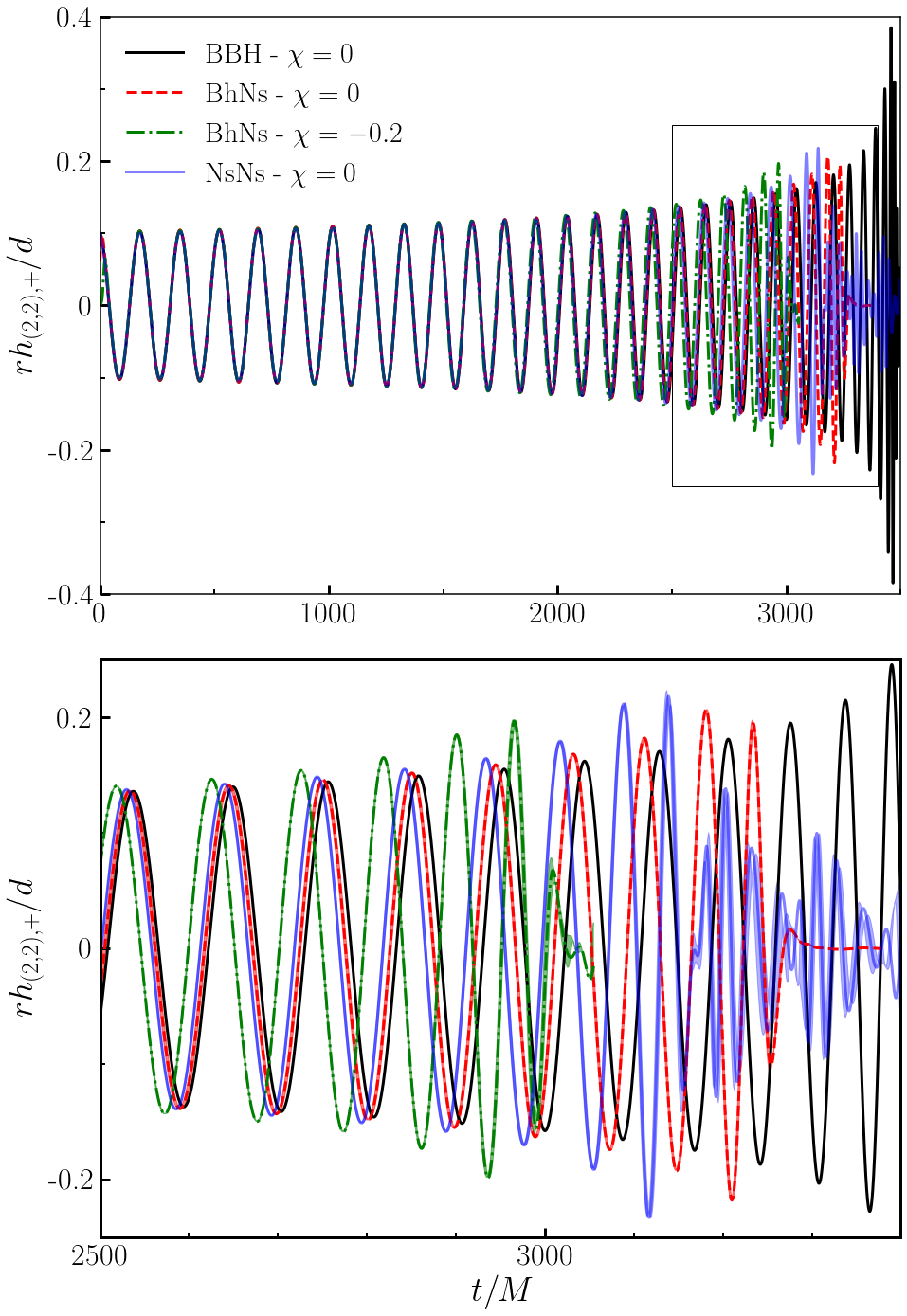}
\caption{Dominant $(2,2)$ mode of the gravitational wave signal for all $q=1$ cases using the $\Gamma 2$ equation of state. The shaded regions in the zoom-in around merger time (bottom panel) lie in between waveforms dephased by the estimated errors from Fig.~\ref{fig:MatchedQ1Err}. The waveform for the binary black hole simulation is assumed to be exact, as errors are significantly smaller for vacuum simulations than for simulations involving neutron stars. All waveforms are aligned through a time and phase shift minimizing the phase difference in the time interval $100<t/M<1100$.}
\label{fig:Q1GW}
\end{center}
\end{figure}

\begin{figure}
\begin{center}
\includegraphics[width=.99\columnwidth]{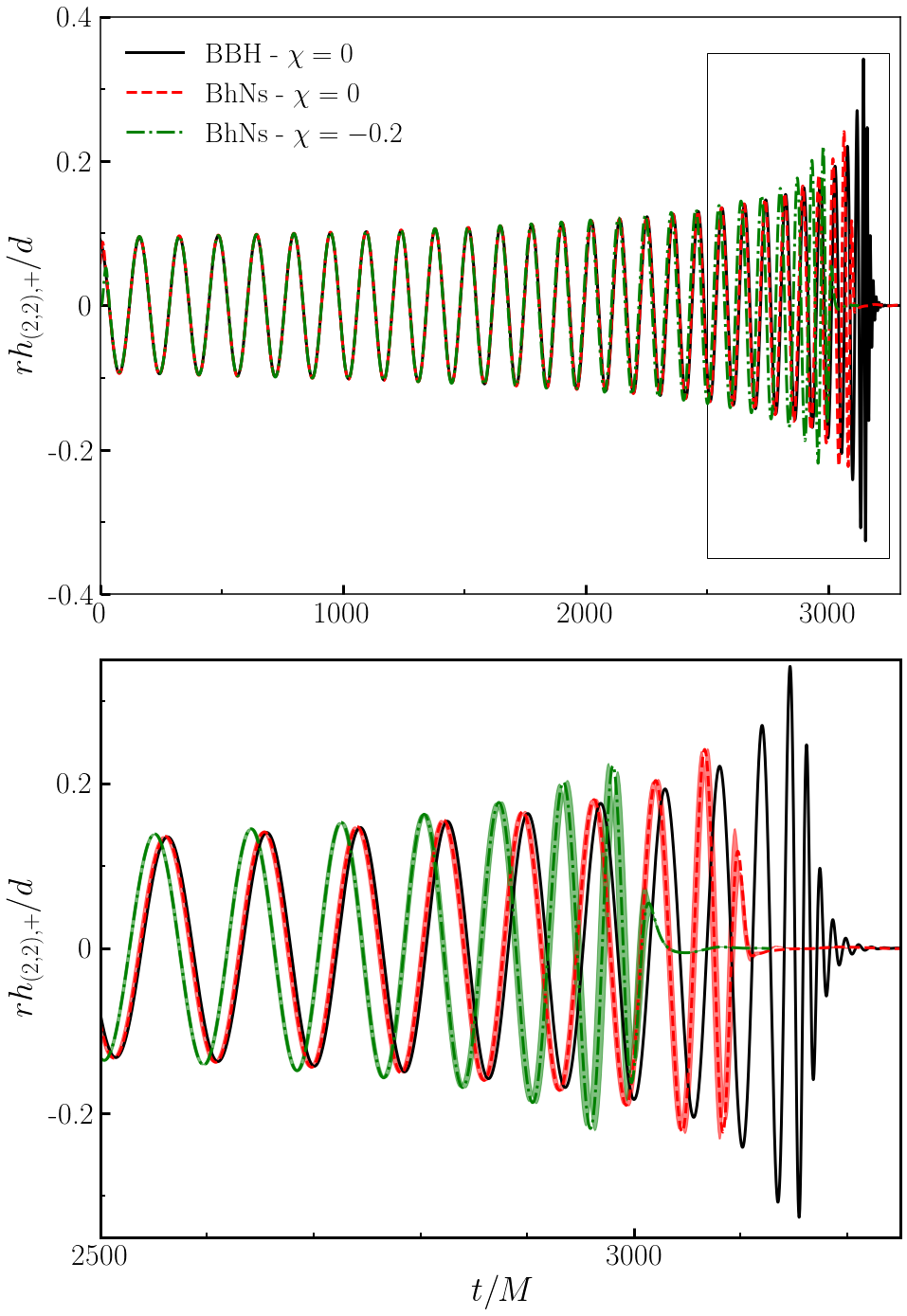}
\caption{Same as Fig.~\ref{fig:Q1GW}, but for the $q=2$ configurations. The errors in the bottom panel are from Fig.~\ref{fig:MatchedQ2Err}.}
\label{fig:Q2GW}
\end{center}
\end{figure}

Numerical waveforms for all the configurations with mass ratio $q=1$ are shown in Fig.~\ref{fig:Q1GW}, and those for $q=2$ in Fig.~\ref{fig:Q2GW}. These figures clearly show the main differences between the three types of binaries. Black hole binaries, lacking tidal dissipation, evolve slower towards merger, and the merger signal itself is followed by the usual exponentially decaying ringdown. Mixed binaries evolve faster, accumulating $(1-2)$rad of dephasing with the black hole binary by the time the neutron star is disrupted by the tidal forces due to the black hole. Tidal disruption cuts off the last $\sim 2$ gravitational wave cycles of the signal for the $q=2$ binary, and as much as $\sim 5$ gravitational wave cycles for the equal-mass system. After disruption, as matter falls into the black hole or forms an accretion disk, there is nearly no gravitational wave emission. Finally, the $q=1$ neutron star binary has, unsurprisingly, tidal effects twice as strong as the $q=1$ mixed binary. The peak of the waveform, as the two neutron stars collide, occurs only slightly earlier than the disruption of the neutron star in the mixed binary system. However, after merger the signal is very different, showing the expected high-frequency oscillations of the remnant. In simulations using more realistic equations of state, these oscillations contain information that can also help constraint the properties of neutron stars~\cite{Bauswein:2014qla,Bauswein:2015vxa,Clark:2015zxa,Takami:2014zpa,Takami:2015}.

From these figures, we can also see that tidal effects are dwarfed by the impact of a high neutron star spin ($\chi=0.2$, antialigned) on the waveforms. The dephasing of the waveform for the spinning mixed binary is $4-10$ times the dephasing of the non-spinning mixed binary. This is consistent with existing results for neutron star binaries indicating that somewhat lower NS spins ($\chi \sim 0.05-0.1$) can have an important impact on gravitational wave signals~\cite{Bernuzzi:2013rza}.

\begin{figure}
\begin{center}
\includegraphics[width=.99\columnwidth]{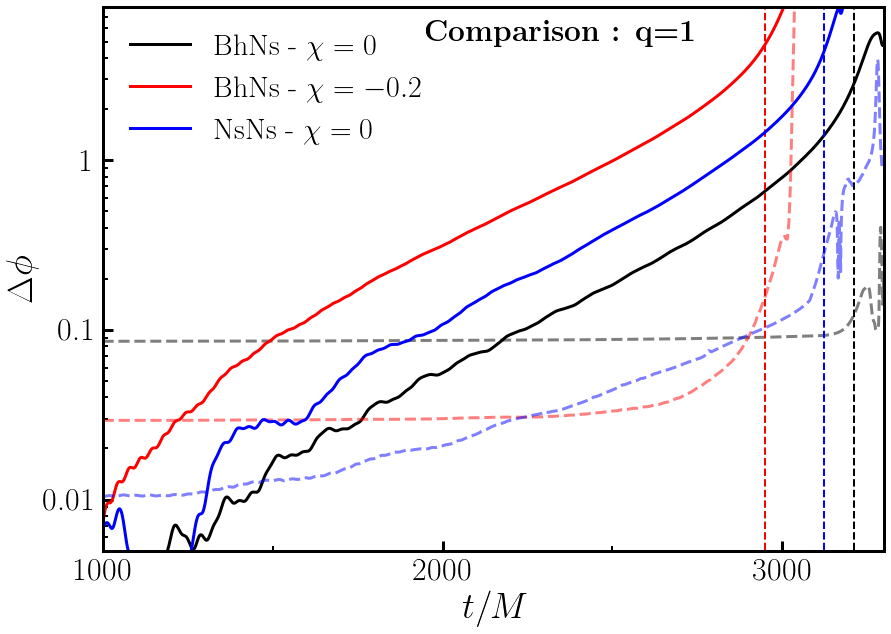}
\caption{Phase difference between the $(2,2)$ modes of the gravitational wave signals of the $q=1$ systems with $\Gamma$-law equation of state, and an equal mass, non-spinning binary black hole waveform.
The waveforms are aligned by applying a time and phase shifts minimizing the phase error in the time interval $100M<t<1100M$ of the non-spinning BHNS system. Dashed curves
show our conservative estimate of the phasing error, aligned over the same time interval, and the vertical lines correspond to the time of peak gravitational wave amplitude for each system. We see that both tidal effects and spin effects are resolved in the simulations, conservatively within a few percents at the peak of the gravitational wave signal ($\sim 10\%$ if using raw numerical error without alignment).}
\label{fig:MatchedQ1Err}
\end{center}
\end{figure}

\begin{figure}
\begin{center}
\includegraphics[width=.99\columnwidth]{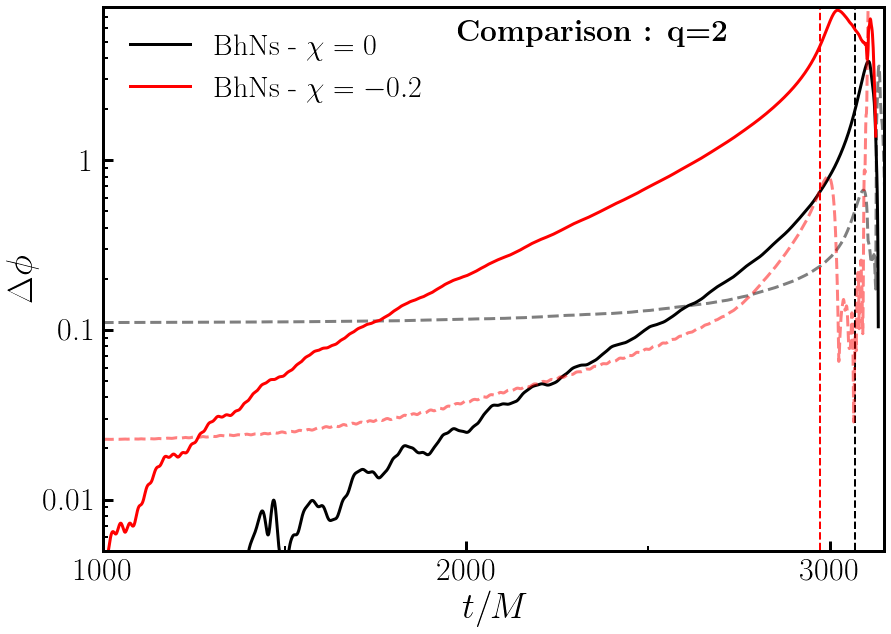}
\caption{Same as Fig.~\ref{fig:MatchedQ1Err}, but for the asymmetric $q=2$ BHNS systems, both compared to a non-spinning $q=2$ binary black hole system. 
As finite-size effects are smaller, and errors larger, we can only guarantee that tidal and spin effects are resolved at the $\sim 25\%$ level at
the peak of the gravitational wave signal (with or without alignment of the waveforms).}
\label{fig:MatchedQ2Err}
\end{center}
\end{figure}

The waveforms presented in Figs.~\ref{fig:Q1GW}-\ref{fig:Q2GW} are aligned by adding small time and phase shifts chosen to minimize phasing errors in the interval $100<t/M<1100$ (more precisely, the root-mean-square of the phasing error sampled every $\Delta t=1M$). To determine how well we resolve differences between black hole, neutron star, and mixed binaries, it is useful to construct error estimates that take into account this matching procedure. We thus repeat the procedure from Sec.~\ref{sec:errors} after aligning waveforms at different resolution / using different order of extrapolation in the same time interval $100<t/M<1100$. The resulting error estimates are shown in Fig.~\ref{fig:MatchedQ1Err} ($q=1$) and Fig.~\ref{fig:MatchedQ2Err} ($q=2$). This alignment procedure nearly uniformly reduces our estimate of the discretization error, but can significantly increase our estimate of the extrapolation error (we do not modify the estimate of the mass loss error). Larger extrapolation errors can occur after the matching procedure because small extrapolation errors in the matching interval lead us to choose a non-zero time-shift between waveforms computed using different orders of extrapolation, which translates into more significant phase errors close to merger. From a numerical point of view, this is not a ``real'' error. We know that we should not apply any time shift between waveforms computed using different extrapolation orders. However, this extrapolation error is meaningful for waveform comparisons, because it corresponds to a very real uncertainty in the matching procedure. Another way to see this is that slightly different phase evolution for waveforms extrapolated using different methods lead to an uncertainty in the frequency of the gravitational wave in the matching interval, thus complicating the alignment of waveforms that do not start from the same initial data.

From Figs.~\ref{fig:MatchedQ1Err}-\ref{fig:MatchedQ2Err}, we gather that our simulations have errors of the order of $(5-10)\%$ [resp. $\sim 25\%$] of the accumulated phase difference due to finite-size effects for $q=1$ [resp. $q=2$] binaries. These results are an important indication of how far our current numerical waveforms can go in constraining analytical waveform models including tidal effects.

\section{Comparison with analytical models}
\label{sec:models}

\begin{figure*}
\begin{center}
\includegraphics[width=.99\textwidth]{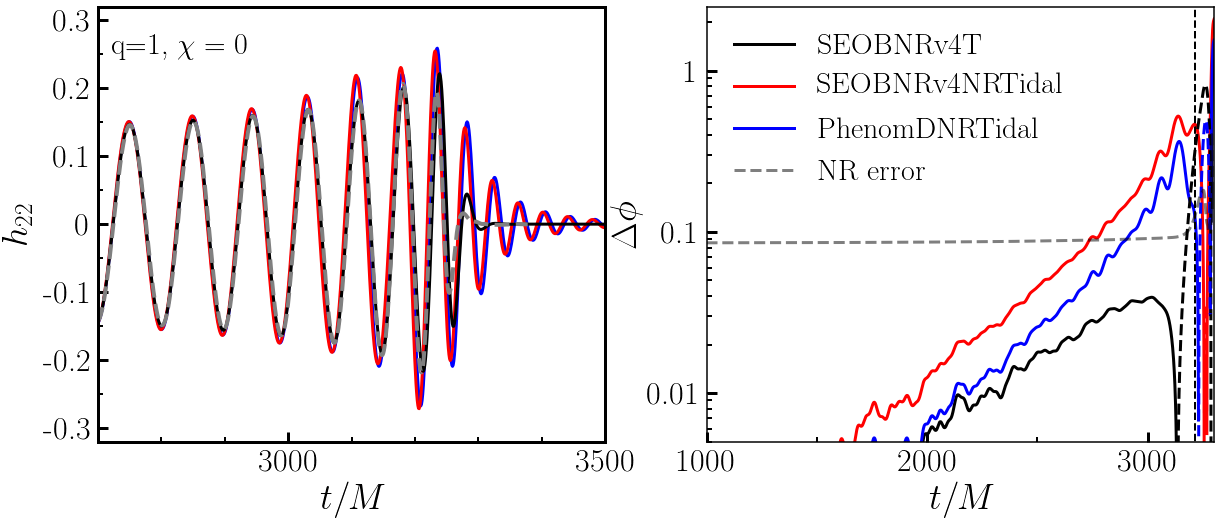}
\includegraphics[width=.99\textwidth]{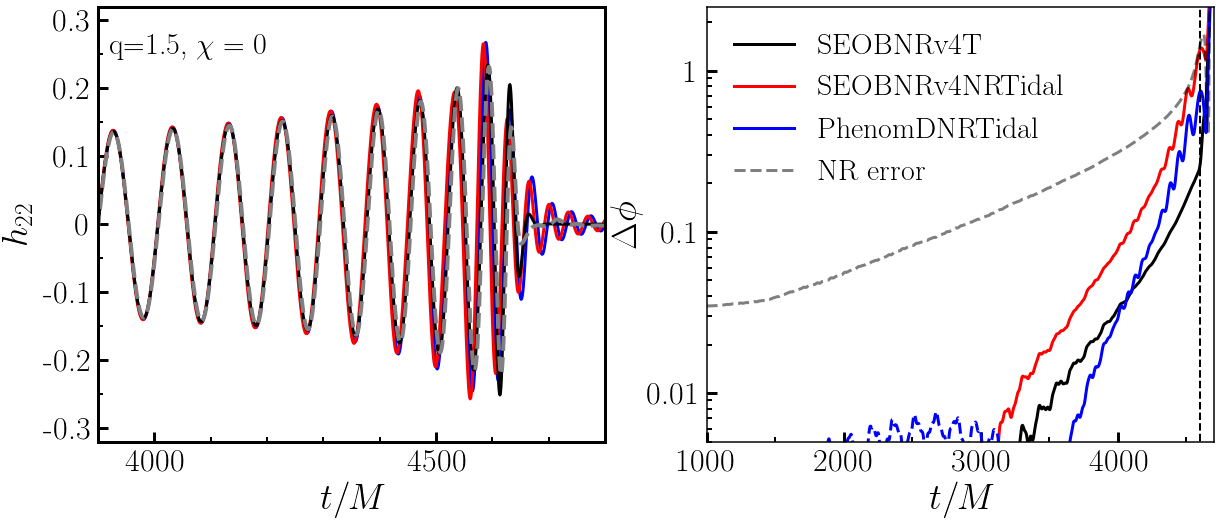}\\
\includegraphics[width=.99\textwidth]{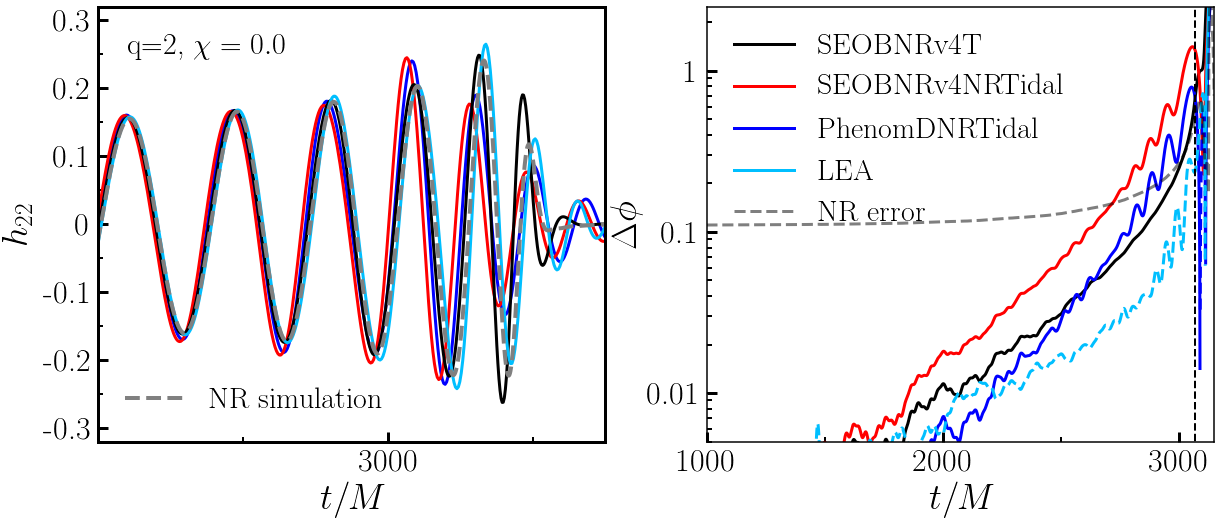}\\
\caption{Comparison between numerical waveforms and analytical models for non-spinning BHNS binaries. For each configuration, the left panel shows the amplitude of the '+' polarization of the dominant $(2,2)$ mode of the gravitational wave signal, zooming in on the region where models and simulations diverge (the gray curves are numerical results, while other curves are model predictions). The right panel shows phase differences between analytical models and the highest resolution numerical waveform at our disposal. In that panel, solid lines denote regions where the analytical model is ahead of the simulation, and dashed lines regions where the simulation is ahead of the model. The dashed vertical line in the right panel corresponds to the peak of the GW signal. LEA is the only model used here that attempts to capture the waveform past that peak.}
\label{fig:ModelsBHNSNoSpin}
\end{center}
\end{figure*}

\begin{figure*}
\begin{center}
\includegraphics[width=.99\textwidth]{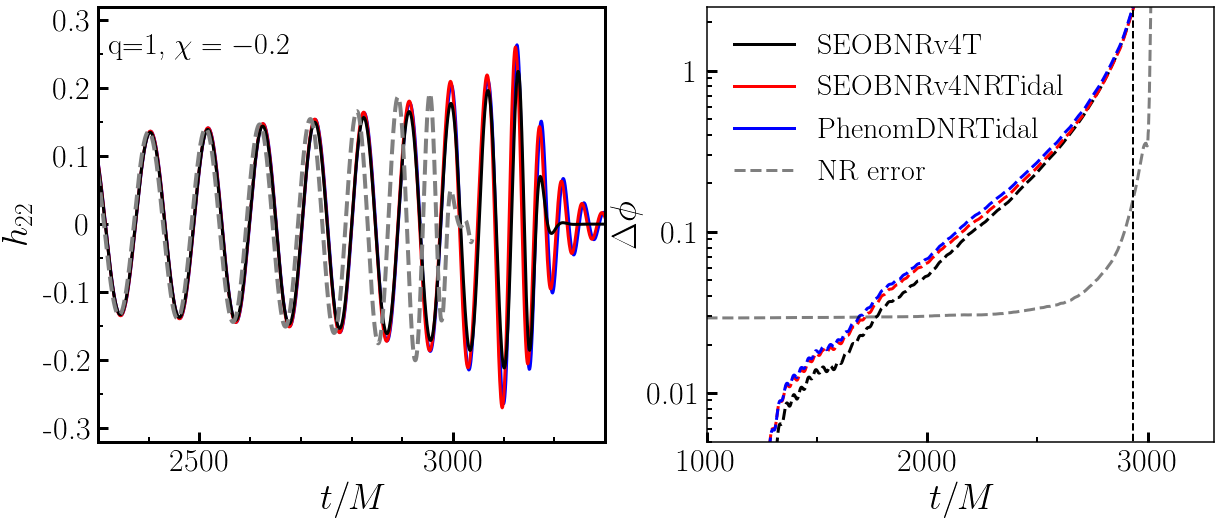}\\
\includegraphics[width=.99\textwidth]{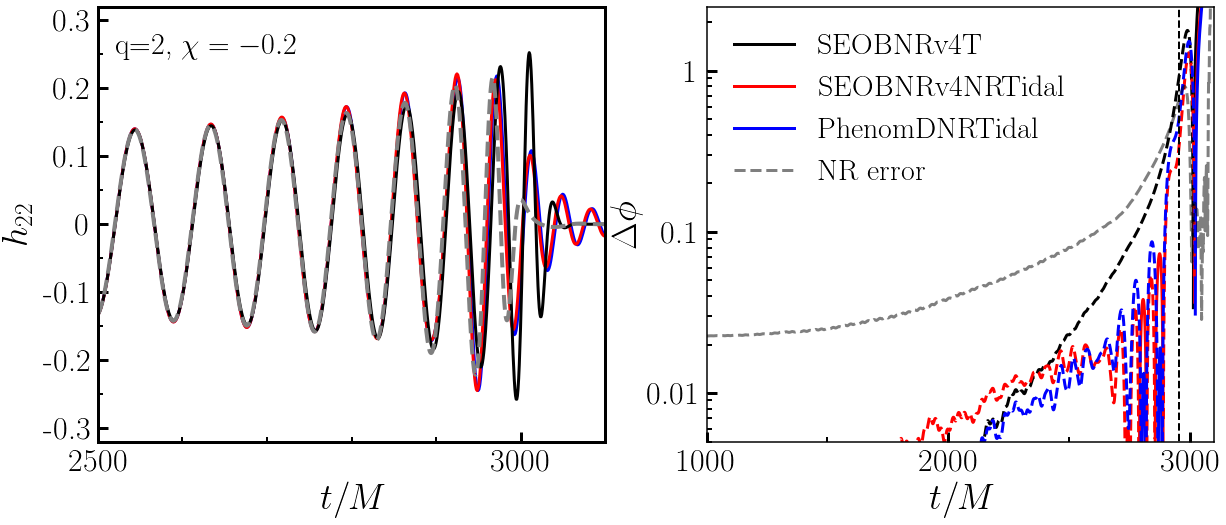}\\
\includegraphics[width=.99\textwidth]{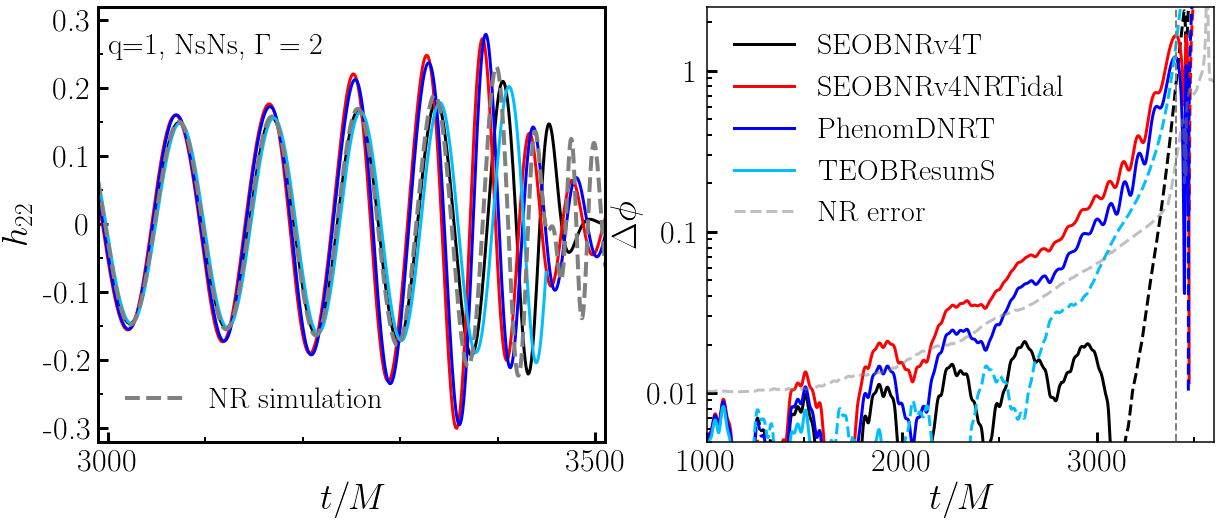}
\caption{Same as Fig.~\ref{fig:ModelsBHNSNoSpin}, but for BHNS systems with spinning neutron stars,
and for the equal mass NSNS system with $\Gamma 2$ equation of state. As before, dashed and solid curves denote phase errors of different signs.}
\label{fig:ModelsBHNSSpin}
\end{center}
\end{figure*}

With error estimates at hand, we can now compare our waveforms with publicly available waveform models. We consider five different models that (aside from one exception) are implemented in the publicly available Ligo Algorithms Library (LAL). They differ in the description of relativistic spinning point masses and/or of matter effects, and are available in LAL under the following names: 
\begin{itemize}
\item {\verb SEOBNRv4T } is a time-domain effective one body (EOB) model that uses the BBH baseline SEOBNRv4~\cite{Bohe:2016gbl}, which is based on the structural inputs developed in Refs.~\cite{Barausse:2009xi, Barausse:2011ys,Taracchini:2013rva,Taracchini:2012} and earlier ideas from Refs.~\cite{Buonanno99,Buonanno00,Barausse:2009aa,Damour2007,Damour2007a,Damour:2014yha}, among others. The naming convention is the following: "S" means that spin effects with fully relativistic test-spin limit are incorporated, "EOB" refers to the modeling approach, "v4" refers to the NR calibration version of the non-precessing model. Matter effects are modeled analytically and {\it dynamically} by including f-mode excitations from the quadrupole and octupole~\cite{Hinderer:2016eia,Steinhoff:2016rfi}, but f-mode excitations neglect the effect of 
the NS spin, which we find important. The spin-induced quadrupole effects are included at leading order\footnote{As described in the internal LIGO Technical Document T1800028}. The SEOBNRv4T model describes both NSNS and BHNS inspirals. 
Once the inspiral evolution meets a stopping criterion (e.g. reaches a peak in orbital frequency or the merger frequency of a NSNS binary as determined from a fit to NR data~\cite{Bernuzzi2015}, or the frequency of the f-mode resonance) the waveform is tapered to zero
\footnote{Although Ref.~\cite{Hinderer:2016eia} developed a non-spinning merger-ringdown model for BHNS binaries, we do not employ it here, but use instead the version of SEOBNRv4T available in LAL, which simply tapers the waveform at the peak of the amplitude.}. 
For the comparisons below, we used quasi-universal relations between NS parameters ~\cite{Yagi:2013ilq,Yagi:2013bca,Chan:2014a} to encapsulate the EOS-dependence in a single parameter $\Lambda$.
\item { \verb SEOBNRv4NRTidal } is a frequency-domain reduced-order-model (ROM) version of the BBH baseline of SEOBNRv4 augmented with tidal effects described by the fit to NR from Ref.~\cite{Dietrich:2017aum}, assuming that the EOS-dependence is characterized only by $\Lambda$, and spin-induced quadrupole effects. The model terminates smoothly beyond the NSNS merger frequency from ~\cite{Bernuzzi2015}. Although the NRTidal model and stopping criteria are tuned to NSNS binaries, waveforms can also be generated for BHNS binaries. 
\item {\verb PhenomDNRTidal } also describes matter effects through the fit to NR from Ref.~\cite{Dietrich:2017aum}. The tidal part is added to a frequency-domain phenomenological (Phenom) BBH baseline model with NR calibration version ``D'' for non-precessing objects from Refs.~\cite{Khan:2015jqa,Husa:2015iqa}, and also earlier work in Refs.~\cite{Ajith:2007qp,Ajith-Babak-Chen-etal:2007b,Ajith2009,Santamaria:2010yb}. The model describes the inspiral phase up to the NS-NS merger frequency~\cite{Bernuzzi2015}, and as SEOBNRv4NRTidal, can also be generated for BHNS binaries.
\item {\verb TEOBResumS } 
is not available in LAL but upon request from the developers. The model is constructed using the EOB formalism but the BBH baseline is built from Refs.~\cite{Damour01c,Damour:024009,Nagar:2011fx,Balmelli:2013zna,Damour:2014sva}, thus it differs from the one used in the SEOBNRv4 model described above (see Ref.~\cite{Barack:2018yly} for a description of the differences). The quadrupole and octupole spin-induced effects are incorporated in a resummed form, and tidal terms are included adiabatically and are enhanced toward merger through a gravitational self-force description~\cite{Damour:2009wj,Bini:2012gu,Bini:2014,Bernuzzi:2014owa,Nagar:2018zoe}. This model is currently restricted to NSNS binaries.
\item  {\verb LEA } 
is an approximate inspiral-merger-ringdown model for matter effects in BHNS binaries including tidal disruption that was developed by Lackey et al.~\cite{Lackey:2013axa}, also assuming that $\Lambda$ suffices to model the EOS-dependence, and is based on numerical simulations. This matter model is implemented on top of the SEOBNRv2~\cite{Taracchini:2013rva} BBH baseline using the frequency-domain ROM version described in Ref.~\cite{Kumar:2016zlj}. Waveforms can only be generated for $q\geq 2$, nonspinning NSs, and BHs with moderate aligned spins. The overlap with our simulations is thus limited to the single case BHNSq2s0.
\end{itemize}

For all configurations, we compare numerical results with model waveforms after aligning the waveforms in time and phase by minimizing the phase difference in the time interval $t/M \in [100,1100]$ of our highest resolution numerical waveform. Numerical errors are estimated taking that matching procedure into account, as in the previous section. 
Results of these comparisons are shown in Fig.~\ref{fig:ModelsBHNSNoSpin} for non-spinning BHNS systems, and in Fig.~\ref{fig:ModelsBHNSSpin} for BHNS systems with spinning neutron stars and for the one NSNS system where simulations are sufficiently accurate to place meaningful constraints on the models. 

We first discuss results excluding the case of an equal mass BHNS merger with a spinning neutron star, as that simulation is a clear outlier in our study. For the other systems, we find that SEOBNRv4T has phase errors small compared with the numerical errors, except occasionally right close to the time of merger. SEOBNRv4NRTidal is outside of our estimated error bars for the most accurate simulations over the last $\sim 500M$ of evolution for the $q=\{1,2\}$ non-spinning BHNS systems, and for about half of the simulation length for the equal-mass NSNS system. The PhenomDNRTidal most often falls in between the two EOB models.  Both PhenomDNRTidal and SEOBNRv4NRTidal tend to {\it overestimate} the strength of tidal effects. The SEOBNRv4T, SEOBNRv4NRTidal, and PhenomDNRTidal do not attempt to model the disruption of the neutron star, and thus disagreements in the amplitude of the GW signal after it reaches its peak are unsurprising. 

The LEA model, whose phase and amplitude were directly calibrated to numerical simulations, is very close to the numerical results for the one case where a comparison is possible: it shows high phase accuracy, and a much better qualitative agreement with the amplitude of the numerical waveform than other models. The TEOBResumS shows reasonable agreement for the amplitude of the NSNS waveform, with phase errors that only become large compared to NR results about 4 cycles before merger (and then it {\it underestimates} the strength of tidal effects).

It is also useful to compare our results with Dietrich et al.~\cite{Dietrich:2018uni}. In that manuscript, the authors find that for NSNS mergers with stiff equations of state and/or spinning neutron stars, SEOBNRv4NRTidal and PhenomDNRTidal perform much better that waveform models based on Post-Newtonian theory (which we do not consider here). For waveforms matched $\sim 3000M$ before merger, Dietrich et al. find phase differences of $\Delta \phi \sim (1-2){\rm rad}$ at merger between these two models and numerical results, with the analytical models merging before the numerical simulations and numerical errors estimated at $0.5-1.5{\rm rad}$. This appears consistent with the results presented here. 

BHNS binaries with spinning neutron stars, particularly the equal-mass system, are generally more poorly modeled than their non-spinning counterparts. While phase accuracy remains good for the $q=2$ system, the amplitude of the waveform at disruption is not well-captured. This is particularly true for SEOBNRv4T: the shutdown of the gravitational wave signal occurs about one cycle too late for that model. For the equal mass system, both phase and amplitude have large errors, and all models miss the shutdown of the gravitational wave signal by 3-4 cycles. This is most likely due to the impact of f-mode excitation close to merger~\cite{Hinderer:2016a}: the f-mode is excited at lower orbital frequencies for counter-rotating neutron stars, and that effect is expected to lead to large errors in the phase of the gravitational wave signal. However, none of the publicly available models include the effect of the spin-induced shift of the f-mode resonance. Our simulations with spinning neutron stars were in fact chosen to maximize the effect of f-mode resonances, and should allow for meaningful tests of analytical models once spin effects are included in the calculation of these resonances. Considering the improved agreement between numerical simulations and SEOBNRv4T observed in~\cite{Hinderer:2016a} when accounting for f-mode excitations, it is likely that taking into account the shift of the f-mode frequency for spinning neutron stars will greatly reduce the disagreement between models and simulations.

Whether current model accuracy is ``sufficient'' for parameter estimation purposes is a more complex question, that we do not directly attempt to address here. The acceptable level of systematic errors in waveform models depends on the signal-to-noise ratio of the source(s), the noise curve of the detectors, and the properties of the merging objects themselves. For GW170817, the tidal deformability still has $\sim 70\%$ relative uncertainty~\cite{GW170817-PE}, and so all models tested in this paper are likely accurate enough to obtain reasonable bounds on that parameter -- a determination that was already reached by the LVC through comparisons of binary parameters recovered using different models~\cite{GW170817-PE}. One possibly important difference to note between the numerical and analytical waveforms, however, is that with the exception of the equal mass system with a rapidly rotating neutron star, analytical models deviate from numerical results by inspiraling {\it faster} than the simulations. This would lead us to {\it underestimate} the tidal parameter $\tilde \Lambda$ when using these models for parameter estimation. Neglecting the shift in the excitation frequency of the f-mode for spinning neutron stars has the opposite effect.

\section{Conclusions}

We present a first SpEC catalogue of NSNS and BHNS binaries. All configurations are simulated at 3 different resolutions, and we provide conservative error estimates for each binary system. The catalogue contains a series of non-spinning BHNS binary mergers of low mass ratios ($q=1-3$), as well as the first numerical waveforms for low-eccentricity BHNS mergers with spinning neutron stars, and 2 equal mass NSNS binary mergers. The majority of these systems (including all of our most accurate simulations) use a simple ideal gas equation of state to represent the neutron star, in order to minimize numerical errors. Those simulations provide $21-33$ GW cycles, and resolve the dephasing due to tidal effects with $\sim(10-25)\%$ relative errors at merger.

Our numerical results are compared to a number of publicly available waveform models. All models show $\lesssim 1{\rm rad}$ accuracy for the phase of the gravitational waveform when models and simulated waveforms are aligned over the first $\sim 1000M$ of the simulation. While this qualitative agreement is very encouraging, some of the modeled waveforms lie noticeably outside of the simulation errors, leaving room for model improvements. Another important result of our study is that using the difference between analytical models as an estimate of the waveform modeling error appears to provide error bars consistent with our simulation results: we do not observe any systematic deviations between the models and the simulations. This is reassuring, as comparing parameter estimate results using different waveform models is one of the methods currently used to assess errors in the measurement of the tidal deformability of neutron stars.

Over the last few orbits, the amplitude of the gravitational wave signal is more poorly modeled than its phase. The merger portion of the waveform does not capture very well (or does not attempt to model) the complex dynamics of a BHNS/NSNS merger. Yet, as for the phase error, the amplitude differences between models appear to provide a good proxy for the modeling error.

The exception to these rules is the equal mass BHNS binary with a rapidly spinning (retrograde) neutron star. For that configuration, systematic differences between models and simulations are clearly measured. More precisely, the numerical simulation predicts a faster inspiral and earlier shut-down of the GW signal than the waveform models. This is expected if, as recently predicted~\cite{Hinderer:2016a}, resonant excitation of the f-mode of the neutron star plays a significant role in the phase evolution of the system close to merger. For counter-rotating neutron stars, the resonance between the f-mode and the orbital motion of the binary shifts to lower frequencies, and more strongly affect the evolution of the system. For non-spinning system, the f-mode frequency is above the merger frequency, and resonant excitation of the neutron star is strongly suppressed. As the only model that explicitly takes into account f-mode excitation in the evolution of the system ignores that frequency shift for spinning neutron stars, it is not surprising that none of the models used in this paper can capture that effect.

All of the simulations presented in this manuscript are now publicly available. We expect that their main use in the future will be for the calibration of improved analytical models, and possibly additional cross-code comparisons.

\acknowledgments
The authors thank Maximiliano Ujevic for producing the initial data for case NSNSq1MS1b, and Jan Steinhoff and the members of the SxS collaboration
for useful discussions and comments throughout this project.
F.F. gratefully acknowledges support from NASA through grant 80NSSC18K0565, and from the NSF through grant PHY-1806278.
TH is grateful for support from the DeltaITP. AW acknowledges support from NWO VIDI and TOP Grants of the Innovational Research Incentives Scheme (Vernieuwingsimpuls) financed by the Netherlands Organization for Scientific Research (NWO)
H.P. gratefully acknowledges support from the NSERC Canada. 
M.D. acknowledges support through NSF Grant PHY-1806207. 
RH gratefully acknowledges support from NSF grants
ACI-1238993, OAC-1550514 and CCF-1551592.
M.B. and L.K. acknowledge support from NSF grant PHY-1606654 at Cornell,
while the authors at Caltech acknowledge support from NSF Grants PHY-170212 and PHY-1708213. 
Authors at both Cornell and Caltech also thank the Sherman Fairchild Foundation for their support.
Computations were performed on the supercomputer Briar\'ee from the Universit\'e de Montr\'eal, 
managed by Calcul Qu\'ebec and Compute Canada. The operation of these supercomputers is funded
by the Canada Foundation for Innovation (CFI), NanoQu\'ebec, RMGA and the Fonds de recherche du Qu\'ebec - Nature et
Technologie (FRQ-NT). This research is part of the Blue Waters sustained-petascale computing project, which is supported by the National Science Foundation (awards OCI-0725070 and ACI-1238993) and the state of Illinois. Blue Waters is a joint effort of the University of Illinois at Urbana-Champaign and its National Center for Supercomputing Applications. This work is also part of the "PRAC Title TBD" PRAC allocation support by the National Science Foundation (award number OCI TBD).
Simulations were also performed on the Zwicky cluster at Caltech, supported by the Sherman
Fairchild Foundation and by NSF award PHY-0960291. 

\bibliographystyle{iopart-num}
\bibliography{References/References.bib}

\end{document}